# Tunable Ferromagnetism and Thermally Induced Spin Flip in Vanadium-doped Tungsten Diselenide Monolayers at Room Temperature

Yen Thi Hai Pham[1,a], Mingzu Liu[2,4,a], Valery Ortiz Jimenez[1], Fu Zhang[3,4], Vijaysankar Kalappattil[1], Zhuohang Yu[2,3], Ke Wang[5], Teague Williams[5], Mauricio Terrones[2,3,4,6*], and Manh-Huong Phan[1*]

[1]Department of Physics, University of South Florida, Tampa, Florida 33620, USA

[2]Department of Physics, The Pennsylvania State University, University Park, PA 16802 USA

[3]Department of Materials Science and Engineering, The Pennsylvania State University, University Park, PA 16802, USA

[4]Center for Two Dimensional and Layered Materials, The Pennsylvania State University, University Park, PA 16802, USA

[5] Materials Research Institute, The Pennsylvania State University, University Park, PA 16802, USA

[6]Department of Chemistry, The Pennsylvania State University, University Park, PA 16802, USA

[a] *Equal contribution to the work.*

**Corresponding authors:** phanm@usf.edu;  mut11@psu.edu

ABSTRACT

The outstanding optoelectronic and valleytronic properties of transition metal dichalcogenides (TMDs) have triggered intense research efforts by the scientific community. An alternative to induce long-range ferromagnetism (FM) in TMDs is by introducing magnetic dopants to form a dilute magnetic semiconductor. Enhancing ferromagnetism in these semiconductors not only represents a key step towards modern TMD-based spintronics, but also enables exploration of new and exciting dimensionality-driven magnetic phenomena. To this end, we show tunable ferromagnetism at room temperature and a thermally induced spin flip (TISF) in monolayers of V-doped $WSe_2$. As vanadium concentrations increase within the $WSe_2$ monolayers the saturation magnetization increases, and it is optimal at ~4at.% vanadium; the highest doping/alloying level ever achieved for V-doped $WSe_2$ monolayers. The TISF occurs at ~175 K and becomes more pronounced upon increasing the temperature towards room temperature. We demonstrate that TISF can be manipulated by changing the vanadium concentration within the $WSe_2$ monolayers. We attribute TISF to the magnetic field and temperature dependent flipping of the nearest W-site magnetic moments that are antiferromagnetically coupled to the V magnetic moments in the ground state. This is fully supported by a recent spin-polarized density functional theory calculation. Our findings pave the way for the development of novel spintronic and valleytronic nanodevices based on atomically thin magnetic semiconductors and stimulate further studies in this rapidly expanding research field of 2D magnetism.

There has been a growing interest in two-dimensional (2D) van der Waals materials due to their potential role in ultralow-power and ultra-compact device applications.[1-5] 2D transition metal dichalcogenides (TMDs) with combined magnetic, electric and optical features are particularly outstanding for the design and fabrication of multifunctional nanodevices.[5-7] Due to the isotropic Heisenberg exchange,[8] the Mermin-Wagner theorem predicts the absence of long-range magnetic order at finite temperatures in 2D materials. TMD monolayers are known to possess no inversion symmetry and their spin-orbit coupling (SOC) results in a significant splitting of the Kramers degeneracy.[5,7] However, the 2017 breakthroughs regarding the discovery of 2D magnetism in monolayer semiconductor crystals (e.g. $CrI_3$) and the observation of layer-dependent magnetic phases open up new paradigms in both fundamental science and device technologies.[9-13] Since these discoveries, a large body of work has been devoted to the realization of long-range ferromagnetic order in monolayer forms of bulk van der Waals materials (e.g. $Cr_2Ge_2Te_6$, $Fe_3GeTe_2$, $Fe_5GeTe_2$, $VI_3$).[14-17] While these systems hold enormous potential for sophisticated magneto-electronics, as well as for the combination of logic and memory for high-performance computing, they are restricted to low temperature, well below 300 K.[9-17] Since the most important technological applications demand magnetic ordering above 300 K, current research is driven by the creation of air-stable room temperature 2D magnets.[18-20]

An alternative strategy to induce long-range ferromagnetism (FM) in TMDs is through the introduction of small quantities of magnetic dopants to form a dilute magnetic semiconductor. Density functional theory (DFT) has predicted tunable ferromagnetism over a wide range of monolayer TMD semiconductors doped with magnetic transition metals, such as V-doped $MoS_2$,[21,22] V-doped $WS_2$,[23] and V-doped $WSe_2$.[24] Calculations by Gao *et al.* have shown an increase in the magnetization and Curie temperature in monolayer V-doped/alloyed

MoS$_2$ with increasing V-concentration up to 9at.%.[22] Duong *et al.* have theoretically and experimentally shown the occurrence of room temperature long-range FM ordering in monolayer V-doped WSe$_2$ for a relatively low (0.1at.%) V-doping level.[24,25] Through a combined study of aberration corrected high-resolution scanning transmission electron microscopy (AC-HRSTEM), magnetization and first-principles calculations, we have recently demonstrated room-temperature FM optimization in monolayer V-doped WS$_2$ at an intermediate V-concentration of ~2at.%, which is consistent with our *ab initio* calculations of a near-ideal random alloy in which magnetism is quenched due to orbital hybridization at too-close vanadium-vanadium distances.[26] While enhancing FM in these semiconductors is the key for developing TMD-based spintronics and quantum computing technology, it has also enabled observation and exploitation of new and exciting dimensionality-driven physical phenomena.[4-7]

In this Letter, we show the enhancement and tunablity of room temperature ferromagnetism in V-doped/alloyed WSe$_2$ monolayer films (optimal around 4at.%V) grown on SiO$_2$ substrates using a single-step powder vaporization method. Such strong magnetic field- and temperature-dependent magnetization enables the observation of a novel magnetic crossover phenomenon associated with a TISF that becomes prominent at room temperature. The origin of TISF is attributed to the magnetic field- and temperature-dependent flipping of the nearest W-site magnetic moments that are antiferromagnetically coupled to the V-magnetic moments. Our results are in good agreement with a recent spin-polarized DFT calculation.[24]

First, the structural properties of pristine and V-doped/alloyed WSe$_2$ (0.5 - 8at.%) monolayer films grown on SiO$_2$ substrates are discussed (Sample synthesis is described in Method, followed the works[26,27]). Atomic resolution high-angle annular dark-field STEM (HAADF-STEM) images indicate the high crystalline quality of the synthesized samples, and show that V replaces W throughout the film (**Fig. 1a-d and Fig. S1**). Vanadium aggregation is quite modest even at 8 at.% V (**Fig. 1d**). Average V-V spacing ($d$) is determined for all V-doped WSe$_2$ samples from the AC-HRSTEM images (**Fig. S2**). $d$ decreases with increasing V concentration (see **Table S1** in Supplementary Information). For example, $d$ decreases from ~4.95 nm for 0.5at.% V to ~1.56 nm for 5at.% V. Such a change in $d$ significantly alters the magnetism of V-doped WSe$_2$ monolayers.

The magnetic properties of the V-doped/alloyed WSe$_2$ monolayer films are characterized by means of magnetic force microscopy (MFM), magnetometry, and anomalous Hall effect (AHE). Unlike pristine WSe$_2$ monolayers that are almost diamagnetic in nature (**Fig. 1e,f**), the vanadium introduction results in long-range ferromagnetic order at room temperature (see **Fig. 1g-k** and **Fig. S3**). These results are also in full agreement with the spin-polarized DFT calculation for a V-doped WSe$_2$ monolayer[24] and the MFM data for a 0.1at.% V-concentration.[25] It is worth mentioning that the magnetic hysteresis (*M-H*) loops for V-doped/alloyed WSe$_2$ monolayers show a distinct ferromagnetic signature at 300 K (**Fig. S3**), and a remarkable increase in saturation magnetization ($M_S$) when increasing the V concentration from 0.5 to 4at.% (**Fig. 2** and **Fig. S4**), thus agreeing well with the DFT calculations.[25,26] This indicates that the FM of V-doped WSe$_2$ monolayers can be further strengthened by optimizing the distribution of dopant-dopant (V-V) neighbor separations. For high V concentrations (e.g. 5at.%), the FM is partially quenched due to orbital hybridization at too-close V-V distances whose magnetic moments are antiferromagnetically coupled (**Fig. S3**). This is fully consistent with the absence of magnetic domains probed by the MFM for the 8at.% V-doped WSe$_2$ monolayer (**Fig. 1l,m**). This trend is also similar to that observed previously for V-doped WS$_2$ monolayers[26]; the FM signal of this system is relatively weaker though. While room temperature FM has been previously reported by the MFM in V-doped WSe$_2$ monolayers at very low V-doping concentrations (0.1-0.3 at.%)[25], here we show enhancement and tunablity of ferromagnetism ($M_S$ and $H_C$) at room temperature in V-doped/alloyed WSe$_2$ monolayers over a wide range of V concentrations (**Fig. S4**). Magnetic measurements were also performed on V-doped WSe$_2$ samples synthesized at different times and the reproducible results are obtained. The room-temperature intrinsic FM nature is independently confirmed by AHE experiments (e.g. for the 4at.% V-concentration, see **Fig. S5**). Such a strong and tunable ferromagnetism enables the full exploration of these 2D magnets and their heterostructures for advanced spintronics and valleytronics.[5,28,29] Our findings also assert the recent DFT predictions of long-range FM tunability in 2D TMD semiconductors through changing the vanadium conceentration.[21-23,26]

From **Figs. 2c,d**, upon lowering the temperature, $H_C$ increases considerably, while $M_S$ decreases gradually in V-doped/alloyed WSe$_2$. The temperature evolution of $M_S$ can be explained with spin-polarized DFT calculations[24] that have recently carried out for 1at.% V-doped WSe$_2$. The calculation results support the presence of an antiferromagnetic (AFM) coupling between V and its nearest W-site (W1) magnetic moments, in addition to the dominant ferromagnetic couplings between neighboring V magnetic moments, and between V and their farther W-site (W2, W3, W4, W5) magnetic moments.[24] A simplified picture that describes coupling of these spins is depicted in **Fig. 2e**. Accordingly, it is suggested that at low temperatures, antiferromagnetic coupling between V-spin up and W1-spin down is strong and remains stable upon the application of an external magnetic field. However, as the temperature is increased, antiferromagnetic coupling becomes weak, and the application of a sufficiently high magnetic field will align the magnetic moment of W1 along the direction of the applied magnetic field. As a result, the total magnetic moment or $M_S$ of V-doped WSe$_2$ monolayers increases with increasing temperature (**Fig. 2c,d**). Another important fact that needs to be accounted for is the AFM coupling of V magnetic moments at too-close V-V distances due to dopant aggregations that often occur in experimentally synthesized V-doped WSe$_2$ monolayers.[26] This AFM V-V coupling becomes competitive with the FM V-V coupling as V concentration increases.[25,26] Upon increasing the temperature, this AFM interaction also becomes weaker, thus increasing the $M_S$ of the sample. The decrease of $H_C$ with increasing temperature is attributed to the increase in thermal fluctuations of the magnetic moments, a common feature also observed previously for monolayer V-doped WS$_2$ films.[26]

Magnetometry measurements on the V-doped/alloyed WSe$_2$ monolayers also revealed the emergence of a new magnetic crossover between the ascending and descending branches of the *M-H* loop at and above a so-called crossover temperature ($T_{cr}$ ~170 - 190 K) (**Fig. 3** and **Fig. S3**). We define the magnetic field values at which the ascending and descending branches of the *M-H* loop cross at a low field and emerge at high field range to be $H_{cr}$ and $H_S$, respectively. These critical fields were observed to depend largely on temperature; $H_{cr}$ decreased while $H_S$ increased with increasing temperature from 175 to 300 K (**Fig. 3b,c**). Over the low temperature region ($T <$ ~175 K), ferromagnetic and antiferromagnetic spins are coherently coupled, and the film reveals a conventional magnetic hysteresis loop (**inset (i)** of **Fig. 3a**). However, at high temperatures ($T >$ 175 K), we can attribute the magnetic crossover of the ascending branch of the *M-H* loop to a thermally weakened AFM V-W1 coupling. At these temperatures, the W1 magnetic moments begin to follow the direction of the V magnetic moments when an external magnetic field exceeds a critical value ($H_{cr}$), which leads to the sharp increase in the magnetization (*M*) (**Fig. 3a** and **its inset (ii)**). For the descending branch of the *M-H* loop, due to the Zeeman and hyperfine field effects, gradual rotation of the W1 magnetic moments for $H < H_S$ occurs instead. The temperature $T_{cr}$ ~175 K is considered to be the crossover temperature that separates the TISF state from the low-temperature spin coherent state (SCS). Paying close attention to the temperature dependence of the difference in saturation magnetization, $\Delta M_S$, between the ascending and descending branches of the *M-H* loop, one observes $\Delta M_S$ increases (**Fig. 3d**) and $H_{cr}$ decreases significantly as the temperature increases from 175 to 300 K (**Fig. 3b,c**). Furthermore, both parameters can be tuned by changing the V concentration (**Fig. S6**). Therefore, these findings highlight a new and exciting perspective for simultaneously thermal and magnetic control of the magnetization state in atomically thin magnetic semiconductors at desirable ambient temperatures for wide-ranging applications in spintronics, ultrafast sensing, and high-speed information storage systems.[3-5] We note that TISF is unique to the V-doped WSe$_2$ monolayer films, as it is absent in magnetically defect-induced monolayer semiconductors such as WSe$_2$ and MoS$_2$,[30,31] as well as in metallic monolayer magnets such as VSe$_2$ and MnSe$_2$.[18,19] The TISF is also absent in V-doped/alloyed WS$_2$ monolayers[26], in which V and W1 magnetic moments are ferromagnetically coupled.[32] Since TISF is present, but small, in bulk single crystals,[32] our observation of a large TISF in monolayer V-doped/alloyed WSe$_2$ points to an intrinsic, dimensionality-enhanced novel TISF phenomenon.

Finally, to probe the nature of the exchange coupling between FM and AFM spins in the V-doped WSe$_2$ monolayers, we have proposed a novel measurement protocol that is typically employed to explore the exchange bias (EB) effect in a FM/AFM binary system.[33] Briefly, the EB effect arises due to exchange coupling between nearby FM and AFM spins at an FM/AFM interface. The protocol includes zero-field-cooled (ZFC) and field-cooled (FC) magnetization *vs.* magnetic field measurements, and thus these were performed on the V-doped WSe$_2$ samples, and the results are displayed in **Fig. S7**. For each FC *M-H* measurement, the sample was cooled down in the presence of a magnetic field ($H_{FC}$ = 5 kOe) from 300 K to 10 K, then the *M-H* loop was measured at 10 K. Relative to a symmetric ZFC *M-H* loop, the FC *M-H* loop shows a small horizontal shift with a distinctly enhanced value of the coercive field ($H_C$ increases from 645 Oe to 698 Oe for 1at.% V-doped $WSe_2$ and from 393 Oe to 412 Oe for 4at.% V-alloyed $WSe_2$), indicating a small EB effect. The EB field of the V-doped $WSe_2$ monolayer films, defined as $H_{EB} = 0.5(H^+ + H^-)$ was found to be ~20 Oe, where $H^+$ and $H^-$ are the coercive fields of the ascending and descending branches of the *M-H* loop, respectively. The dynamics of the FM/AFM spins at the irreversible/reversible interface are examined through the training effect of the coercive field ($H_C^{FC}$) obtained from FC *M-H* measurements. The sample was also cooled in $H_{FC}$ = 5 kOe ($H_{FC} > H_S$ ~ 3.5 kOe) from 300 K to 10 K, where the magnetic field was cycled a number of times (**Fig. 4a-c**). As one can clearly observe from **Fig. 4c,d**, $H_C^{FC}$ retains almost unchanged for the first to third cycles of the magnetic field but decreases strongly with successive cycling. The decrease in $H_C^{FC}$ is due to rearrangement of spins occurring at the AFM/FM interfaces. The $H_C(n)$ dependence, where *n* is the number of cycles, for the V-doped $WSe_2$ samples are fit using the equation $H_C = H_C^1 + Bn + Cn^2$, where $H_C^1$ is the coercive field for *n* = 1, and B and C are constants. The $H_C(n)$ dependence is remarkably different from previously reported observations of AFM/FM systems.[33] This can be attributed to a topological arrangement of the V/W spins in monolayer V-$WSe_2$ (**Fig. 2e**), in combination with low dimensionality effects.

In summary, we have demonstrated the enhancement and tunability of room temperature ferromagnetism in V-doped $WSe_2$ monolayers at 0.5 to 5at.%V concentrations. An intriguing and novel thermally induced spin flipping phenomenon has also been observed, which appears to be an intrinsic property of this 2D material. The novel magnetic switching, in conjunction with tunable ferromagnetism, now enables the direct magnetic control of valley splitting in V-$WSe_2$ monolayers at room temperature[28]; an arrangement in stark contrast to the interfacial magnetic exchange field induced by a ferromagnetic substrate (e.g. EuS, CrI$_3$) whose Curie temperature is limited to low temperature (< 100 K).[29,34] Our findings provide an approach for the development of applications in van der Waals spintronics, valleytronics and quantum computing devices.

## Methods

**Samples synthesis**

In a typical process, a deionized water solution of ammonium metatungstate ((NH$_4$)$_6$H$_2$W$_{12}$O$_{40}$), vanadium oxide sulfate (VO[SO$_4$], only for V-doped WSe$_2$ samples), and sodium cholate (C$_{24}$H$_{39}$NaO$_5$) was spin-coated onto SiO$_2$/Si substrates. The substrates were then placed in a quartz tube, along with an alumina boat containing selenium powder located upstream. With a mixture of Ar/H$_2$ (10% H$_2$) gas flushing at 80 sccm, the quartz tube was heated in a furnace at 825 °C for 10 min. After the reaction, the system was left cooling down naturally to room temperature.

**Materials characterization**

The aforementioned pristine and V-doped monolayer WSe$_2$ samples were transferred onto Quantifoil R 2/1 200 mesh TEM grids. HAADF-STEM was then performed on a FEI Titan G$^2$ 60 - 300 kV microscope with a cold field-emission electron source. The microscope was operating at 80 kV with double spherical aberration correction to obtain a high resolution of ~ 0.7 Å. A HAADF detector (collection angle 42 - 244 mrad, camera length 115 mm, beam current 45 pA, beam convergence 30 mrad) was used to acquire Z-contrast STEM imaging.

**Magnetic measurements**

*Magnetic Force Microscopy*

The pristine and V-doped WSe$_2$ samples were magnetized with a strong magnet for approximately 10s before the scans. Both AFM and MFM images were obtained on a Bruker Icon AFM system, in tapping mode for sample surface topography and lift mode for magnetic domain structure observation. A Bruker Sb-doped Si magnetic probe (225 × 35 × 2.8 μm$^3$, with tip radius ~ 30 nm), was used for the measurements. The force constant and the resonant frequency of the tip were approximately 3 N/m and 75 kHz, respectively.

*Vibrating Sample Magnetometry*

Temperature- and magnetic field-dependent magnetization measurements were carried out in a Physical Property Measurement System (PPMS) from Quantum Design with a vibrating sample magnetometer (VSM) magnetometer over a temperature range of 2 – 350 K and fields up to 9 T. To exclude unwanted effects on the magnetization versus magnetic field (*M-H*) loops that can arise from subtracting diamagnetic and paramagnetic backgrounds, we present the as-measured *M-H* loops at all measured temperatures and deduces the saturation magnetization ($M_S$) and coercive field ($H_C$) directly from these loops.

*Anomalous Hall Effect*

The 4%at. V-doped WSe$_2$ monolayer film was etched into the shape of a Hall bar device using ion milling. Etching was done using a shadow mask which was fabricated lithographically. Electrical contacts were made using indium pressed dot and Au wires. An alternating current of 0.1 mA was applied for detecting the anomalous Hall Effect (AHE). A measurement configuration is shown in Fig. S5, and magnetic field (*H*) was applied perpendicular to the sample surface. AHE measurements were performed at 300 K inside the PPMS. The anomalous

Hall resistance (AHR) data was obtained after subtracting the linear contribution from the ordinary Hall effect (OHE).

## Acknowledgments

Work at USF was supported by the U.S. Department of Energy, Office of Basic Energy Sciences, Division of Materials Sciences and Engineering under Award No. DE-FG02-07ER46438 and the VISCOSTONE USA under Award No. 1253113200. Work at PSU was supported by the Air Force Office of Scientific Research (AFOSR) through grant No. FA9550-18-1-0072 and the NSF-IUCRC Center for Atomically Thin Multifunctional Coatings (ATOMIC). The authors acknowledge Prof. Dinh Loc Duong of Sungkyunkwan University for useful discussions, and Prof. Felipe Cervantes-Sodi of Ibero-American University for preliminary synthetic studies.

## Author Contributions

M.L., Z.Y., and F.Z. conducted the synthesis experiments. M.L., K.W., and T.W. conducted the TEM/MFM characterizations. Y.H.T.N., V.O.J., V.K., and M.H.P. performed magnetic measurements and analyzed the magnetic data. V.K. measured and analyzed the AHE data. M.H.P. and M.T. initiated the concept and supervised the whole work. All authors have participated in writing the manuscript and given an approval to the final manuscript.

## Competing interests

The authors declare no competing financial interest.

**Figure Captions**

**Fig. 1** Atomic resolution HAADF-STEM images of pristine (**a**) and V-doped WSe$_2$ monolayers at 0.5at.% (**b**), 5at.% (**c**) and 8at.% (**d**) vanadium. The black dots indicate the presence of V atoms that replace W ones. AFM and MFM images of the pristine (**e,f**), 1at.% V-doped (**g,h**), 4at.% V-doped (**i,k**), and 8at.% V-doped (**l,m**) samples, respectively. While the pristine WSe$_2$ sample shows no magnetic domains (**f**), the 1 at.% and 4at.% V-doped WSe$_2$ samples show noticeable changes in magnetic domain (**h,k**), when the samples are magnetized with the same magnetic field (0.5T). The change in magnetic domain becomes unobvious in the 8at.% V-doped WSe$_2$ sample (**m**), indicating a strong suppression of the FM due to orbital hybridization at too-close vanadium-vanadium spacings.

**Fig. 2** Magnetic hysteresis loops taken at 300 and 10 K for V-doped WSe$_2$ monolayers at 1at.% (**a**) and 4at.% (**b**) vanadium. Temperature dependence of saturation magnetization ($M_S$) and coercive field ($H_C$) for the 1at.% (**c**) and 4at.% (**d**) V-doped samples. (**e**) Directions of V and W spins in the nearest and distant regions are shown by arrows; While the nearest W-site (W1) spin couples antiferromagnetically with the V-spin, the distant W-site (W2, W3, W4, W5) spins couple ferromagnetically with the V-spin. The height of the arrow indicates the magnitude of the magnetic moment.

**Fig. 3** (**a**) Magnetic hysteresis loops indicating the presence of a magnetic crossover at high temperature (300 K) and its disappearance at low temperature (100 K) for the 1at.% V-doped WSe$_2$ sample. Critical fields $H_{cr}$ and $H_S$ and difference in saturation magnetization $\Delta M_S$ are shown; Temperature dependence of $H_{cr}$ and $H_S$ for the (**b**) 1at.% and (**c**) 4at.% V-doped WSe$_2$ samples. (**d**) Temperature dependence of $\Delta M_S$ for the 1at.% and 4at.% V-doped WSe$_2$ samples.

**Fig. 4** Field-cooled (FC) magnetic hysteresis (*M-H*) loops for eight successive measurement times and coercive field ($H_C$) as a function of measurement time (*n*) for the (**a,c**) 1at.% and (**b,d**) 4at.% V-doped WSe$_2$ samples. For *FC M-H* measurement, a magnetic field of 5 kOe was applied while cooling down the sample to 10 K from 300 K, and the *M-H* curve was taken at 10 K. The $H_C(n)$ dependence has been fitted using the equation $H_C = H_C^1 + Bn + Cn^2$, where $H_C^1$ is the coercive field for *n* = 1, and B and C are the constants.

**Figure 1**

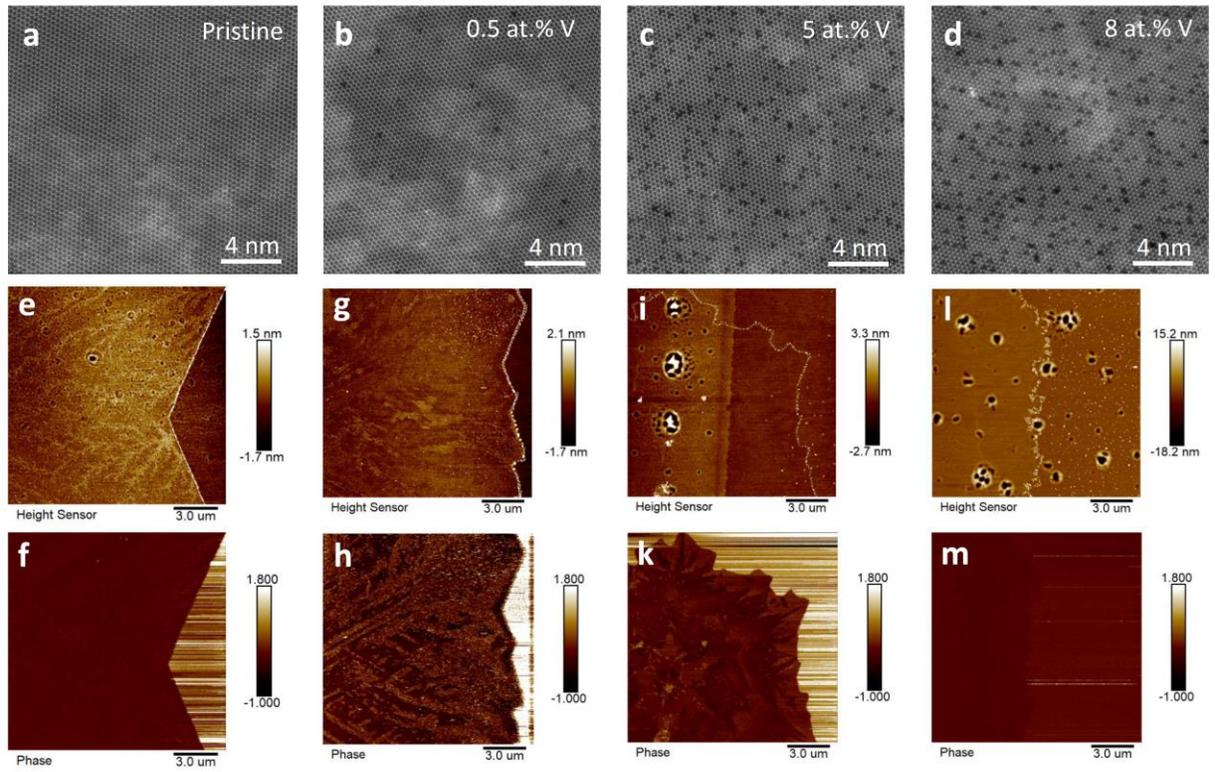

**Figure 2**

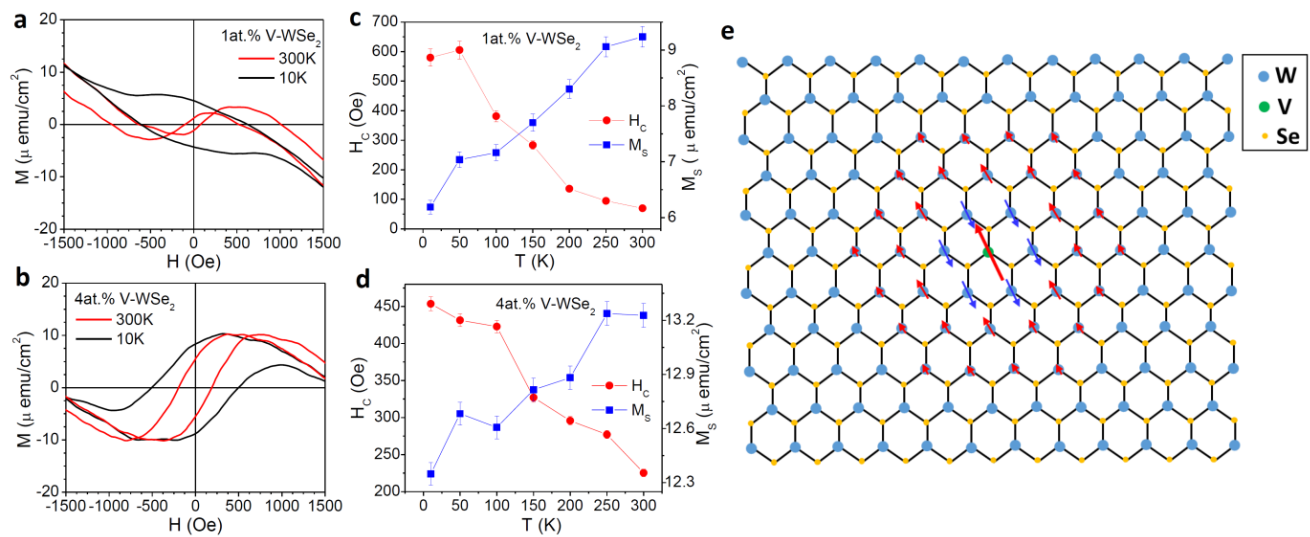

**Figure 3**

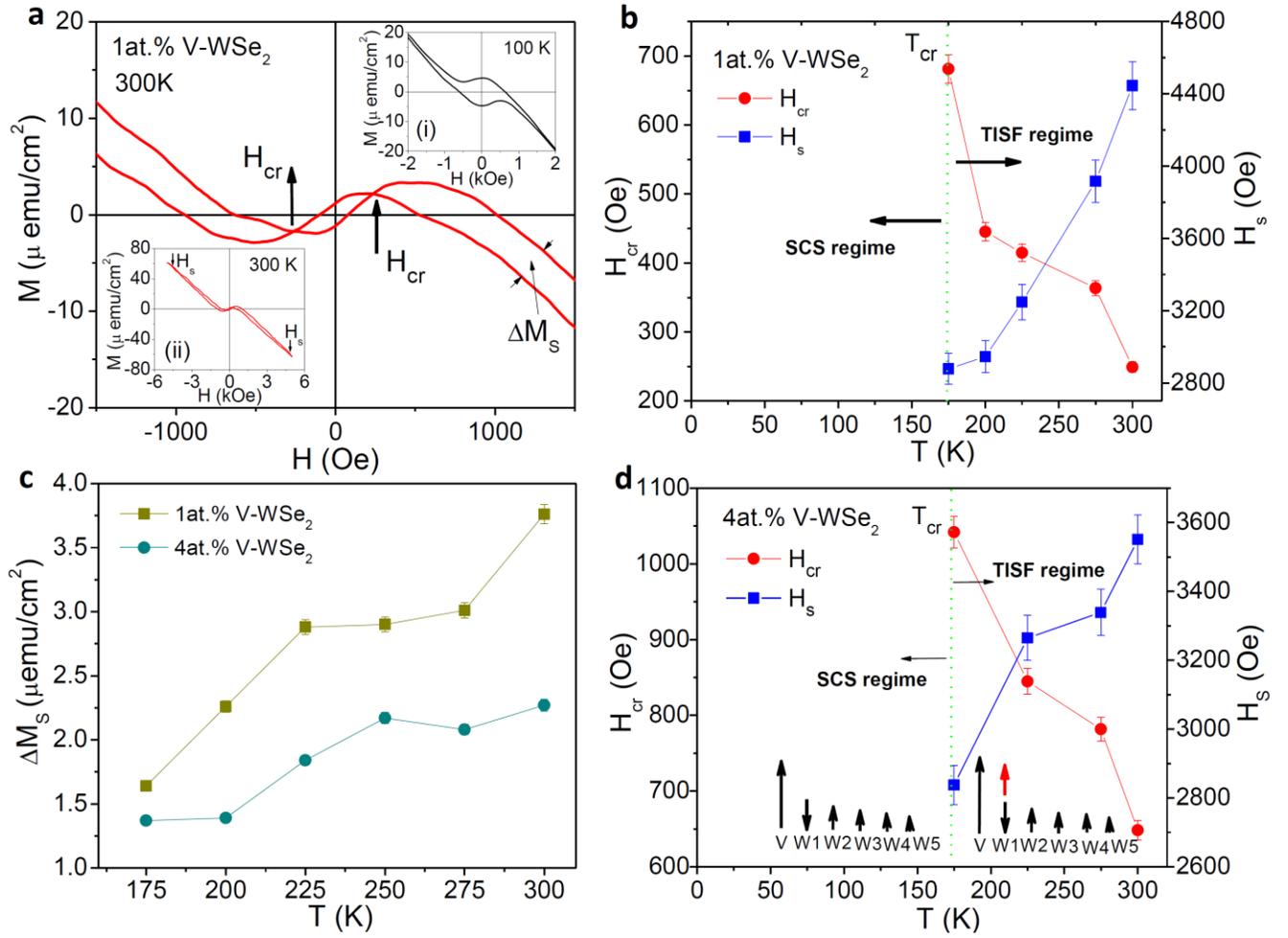

**Figure 4**

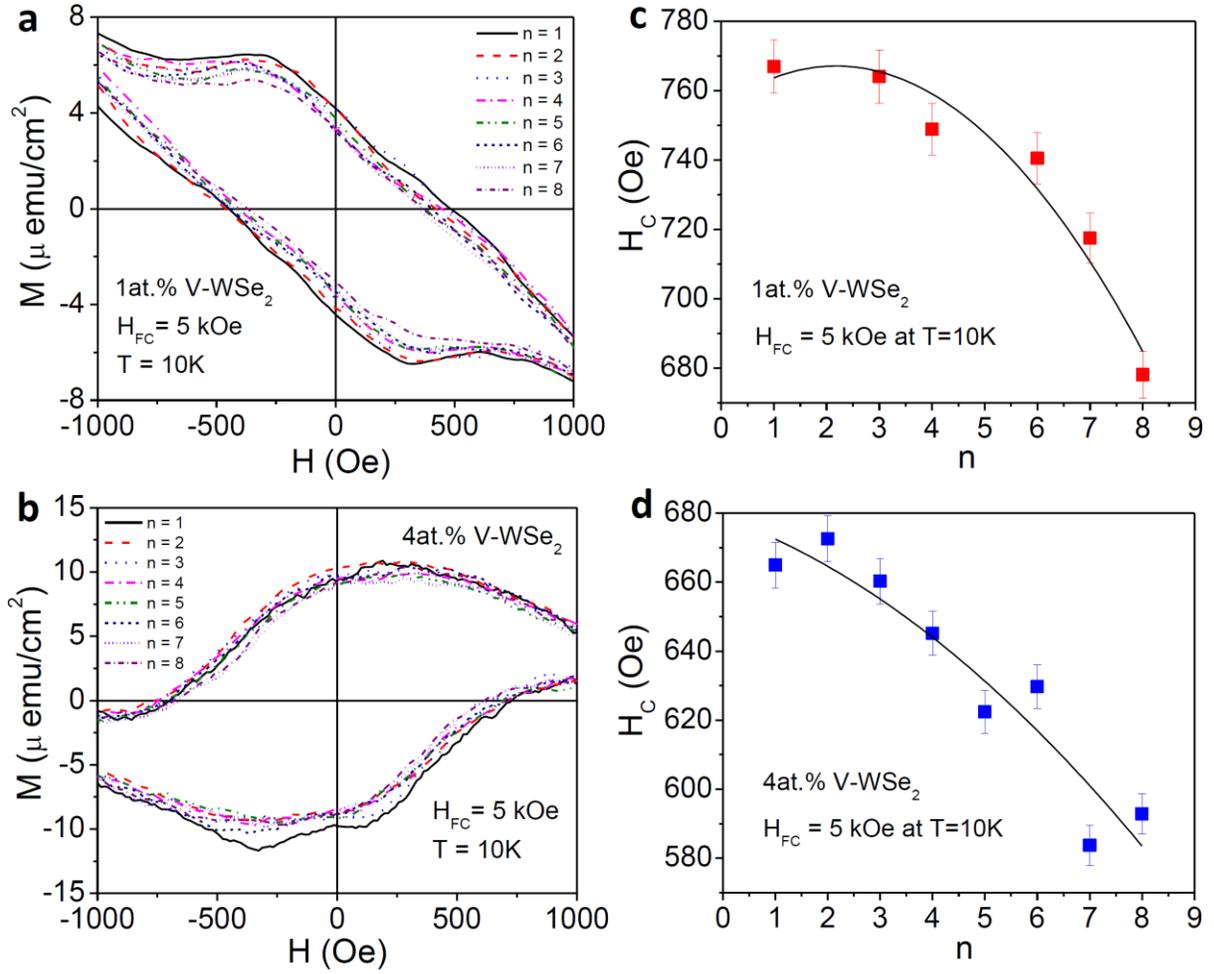
22

Supplementary Information

# Tunable Ferromagnetism and Thermally Induced Spin Flip in Vanadium-doped Tungsten Diselenide Monolayers at Room Temperature

Yen Thi Hai Pham[1,a], Mingzu Liu[2,a], Valery Ortiz Jimenez[1], Fu Zhang[3,4], Vijaysankar Kalappattil[1], Zhuohang Yu[2,3], Ke Wang[5], Teague Williams[5], Mauricio Terrones[2,3,4,6*], and Manh-Huong Phan[1*]

[1]Department of Physics, University of South Florida, Tampa, Florida 33620, USA

[2]Department of Physics, The Pennsylvania State University, University Park, PA 16802 USA

[3]Department of Materials Science and Engineering, The Pennsylvania State University, University Park, PA 16802, USA

[4]Center for Two Dimensional and Layered Materials, The Pennsylvania State University, University Park, PA 16802, USA

[5]Materials Research Institute, The Pennsylvania State University, University Park, PA 16802, USA

[6]Department of Chemistry, The Pennsylvania State University, University Park, PA 16802, USA

[a] *Equal contribution to the work.*

**Corresponding authors:** phanm@usf.edu, mut11@psu.edu

**Figure S1.** Composition gradient on one V-doped WSe$_2$ monolayer flake.

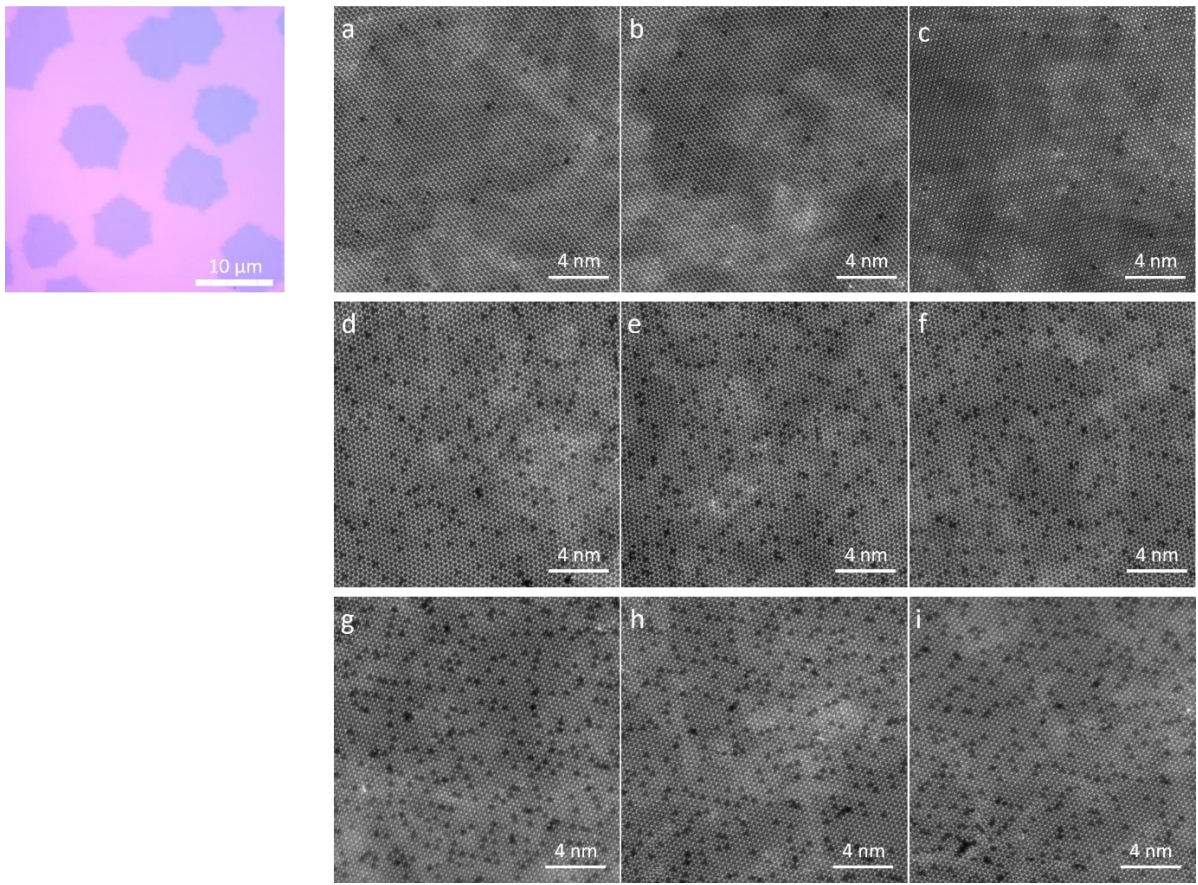

The solution-based CVD as-grown V-doped WSe$_2$ monolayer flakes have near-hexagonal irregular shapes, as shown in **Fig. S1** (left). It is observed that the doping level of vanadium has negligible variance among the flake from center to edge. We randomly chose multiple positions on one flake of different doping levels (**Fig. S1 a-c**: 0.5 at.%; **d-f**: 5 at.%; **g-i**: 8 at.%) and performed STEM measurements. It can be seen that the doping concentration is highly uniform across the flake.

**Figure S2.** STEM images of the V-doped $WSe_2$ monolayer films.

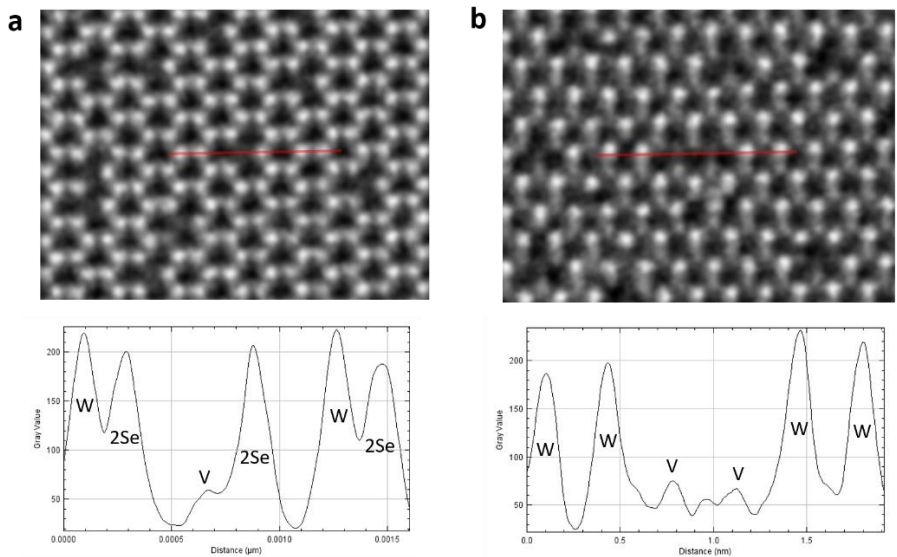

As can be seen from the intensity line profile taken on the experimental STEM image of the 5 at.% V-doped $WSe_2$ sample (**Fig. S2a**), the dopant atom of vanadium is replacing W in the lattice. At higher doping levels of vanadium (8 at.%), the dopant atoms tend to aggregate and form local $VSe_2$ domains, thus achieving alloying between $WSe_2$ and $VSe_2$ (**Fig. S2b**).

**Figure S3.** Magnetization versus magnetic field curves taken at 300 K for pristine and V-doped WSe$_2$ monolayer samples.

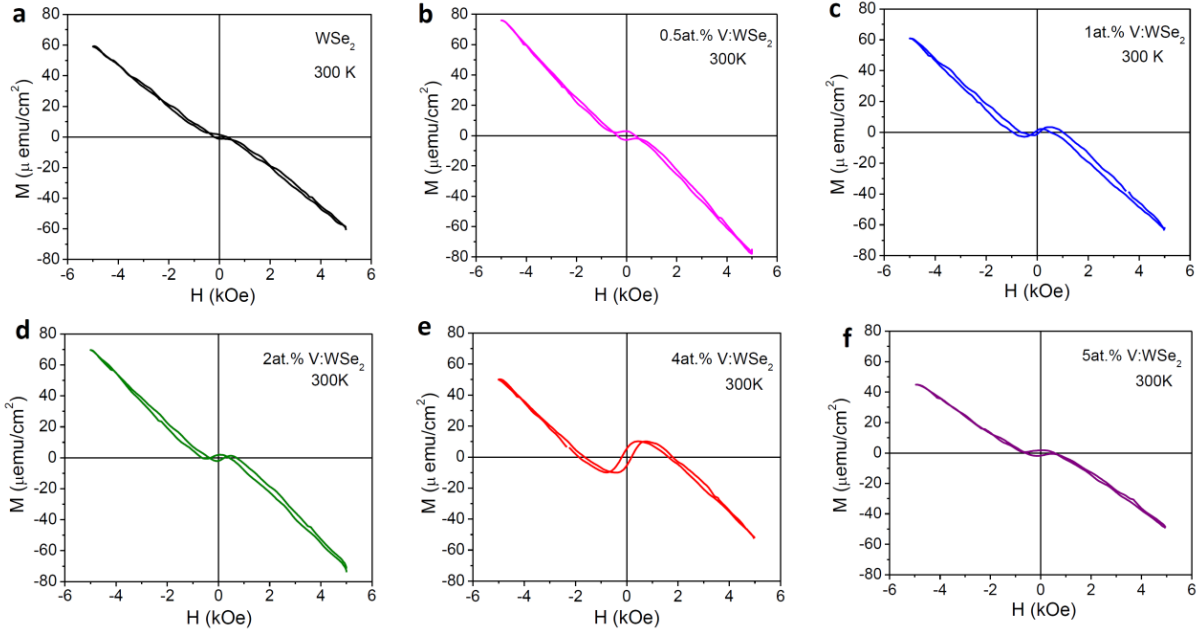

It is observed that while WSe$_2$ is mostly diamagnetic (**Fig. S3a**), vanadium doping results in the emergence of ferromagnetic order in the V-doped WSe$_2$ monolayers. The V-doped WSe$_2$ samples show a magnetic crossover associated with the thermally induced spin fliping of the nearest W-site (W1) magnetic moments that are antiferromagnetically coupled to the V magnetic moments.

**Figure S4.** Saturation magnetization ($M_S$) and coercive field ($H_c$) vary as functions of the V-doping concentration at room temperature.

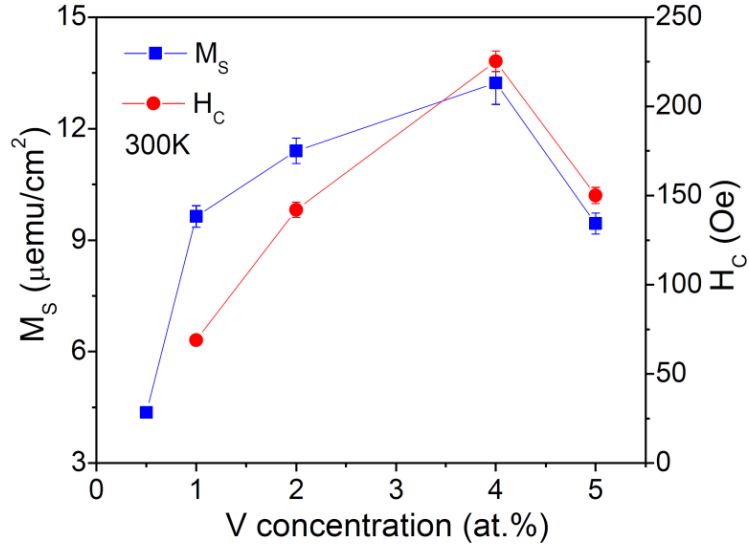

It is observed that the saturation magnetization ($M_S$) and coercive field ($H_C$) increase as V concentration increases from 0.5 to 4 at.%, reache a maximum at ~4 at.%, and then decrease for higher V-doping concentrations (e.g. 5at.%). This result indicates that the ferromagnetism of V-doped WSe$_2$ monolayers can be further strengthened by optimizing the distribution of dopant-dopant (V-V) neighbor separations.

**Figure S5.** Anomalous Hall effect observed for the 4at.% V-doped WSe$_2$ monolayer film

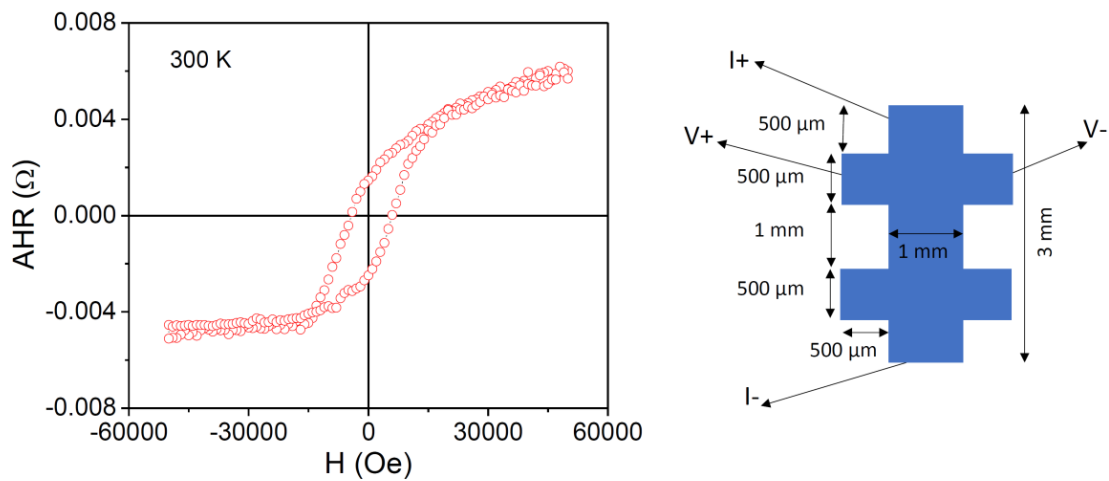

The 4at.% V-doped WSe$_2$ monolayer film shows a hysteretic loop at 300 K (**Fig. S5**, left panel), indicating an intrinsic ferromagnetic characteristic of the material at room temperature. This result is fully consistent with the *M-H* data recorded by the PPMS. A Hall bar shape with a measurement configuration is shown in **Fig. S5** (right panel).

**Figure S6.** $\Delta M_S$ and $H_{cr}$ vary as functions of the V-doping concentration.

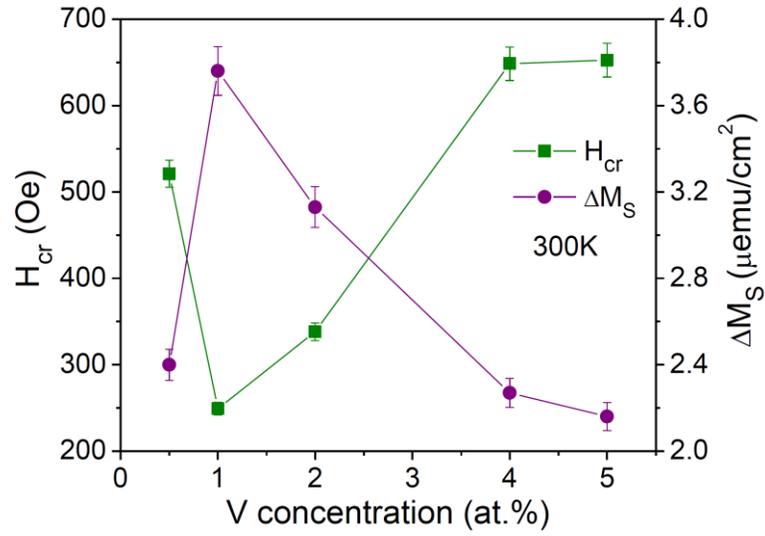

It is observed that the difference in saturation magnetization, $\Delta M_S$, between the ascending and descending branches of the *M-H* loop is maximal and the crossover field ($H_{cr}$) is minimal around 1at.%V. As the V concentration increases from 1 to 5at.%V, $\Delta M_S$ decreases and $H_{cr}$ increases strongly until they reach saturation at 5at.%.

**Figure S7.** Magnetic hysteresis loops (*M-H*) taken at 10 K in zero-field-cooled (ZFC) and field-cooled (FC) regimes for 1 and 4at.% V-doped WSe$_2$ monolayer samples.

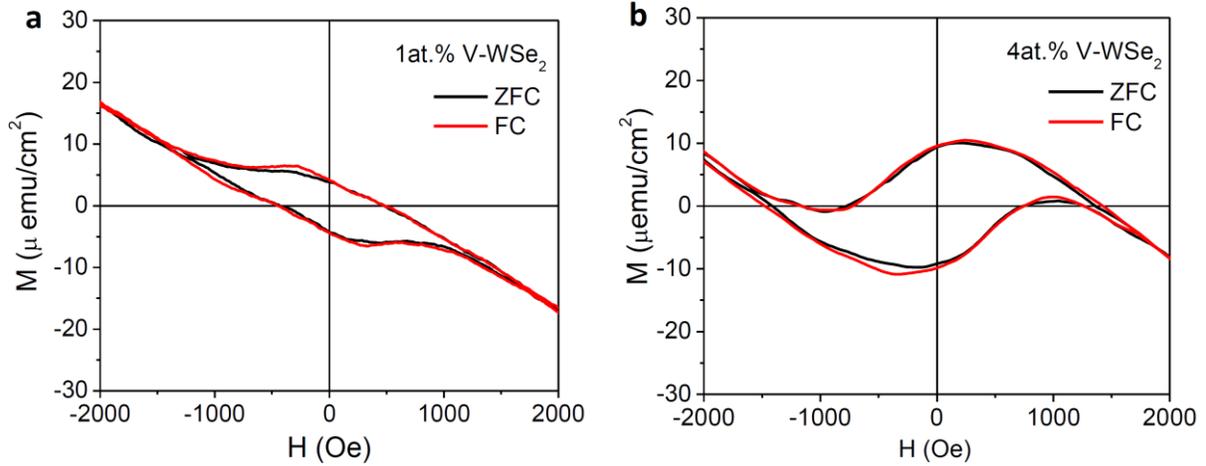

For ZFC measurement, the sample was cooled to 10 K from 300 K in the absence of a magnetic field, and the *M-H* loop was then measured at 10 K. For FC measurement, the sample was cooled to 10 K from 300 K in the presence of a magnetic field (5 kOe), and the *M-H* loop was then taken at 10 K. We have observed a small horizontal shift of the FC *M-H* loop with an enhanced coercivity, indicating the presence of a small exchange bias (EB) effect.

**Table S1.** Average V-V spacings determined from the HR-TEM images of the V-doped WSe$_2$ monolayer film samples.

| Doping concentration | | average dopant spacing (nm) |
|---|---|---|
| (at.%) | (cm$^{-2}$) | |
| 0.5 | 4.1×10$^{12}$ | 4.95 |
| 5 | 4.1×10$^{13}$ | 1.56 |
| 8 | 6.5×10$^{13}$ | 1.24 |

The average V-V spacings $d$ are obtained from $d = n^{-1/2}$, where $n$ is the doping concentration of V in cm$^{-2}$ converted from at.%. This is based on the observation from our STEM images, that the V dopant atoms are uniformly distributed in the WSe$_2$ lattice.